\documentclass[prd,showpacs,onecolumn]{revtex4}
\usepackage{psfrag,graphics,dcolumn,bm}

\begin{document}

\title{Theoretical uncertainty in baryon oscillations}

\author{Daniel Eisenstein}
\affiliation{Steward Observatory, University of Arizona, Tucson, AZ 85721}
\author{Martin White}
\affiliation{Departments of Physics and Astronomy, University of California,
Berkeley, CA 94720}
\date{\today}

\begin{abstract}
We discuss the systematic uncertainties in the recovery of dark energy
properties from the use of baryon acoustic oscillations as a standard ruler.
We demonstrate that while unknown relativistic components in the universe
prior to recombination would alter the sound speed, the
inferences for dark energy from low-redshift surveys are unchanged so
long as the microwave background anisotropies can measure the redshift of
matter-radiation equality, which they can do to sufficient accuracy.
The mismeasurement of the radiation and matter densities themselves
(as opposed to their ratio) would manifest as an incorrect prediction for
the Hubble constant at low redshift.  In addition, these anomalies do
produce subtle but detectable features in the microwave anisotropies.
\end{abstract}


\pacs{98.80.Es,95.85.Bh,98.35.Ce,98.70.Vc \hfill}

\maketitle

\section{Introduction}

In standard cosmology, the acoustic oscillations imprinted in the matter 
power spectrum at recombination have a length scale that can be 
accurately calculated based on measurements of the CMB anisotropy 
power spectrum \cite{PeeYu,BE84,Hol,EHT,MWP}.
It should then be possible to measure this ``standard ruler'' scale 
at low redshifts,
for example in large galaxy redshift surveys, and thereby constrain the
matter and energy content of the universe
\cite{Eis03,BlaGla,Lin,HuHai,SeoEis}.
However, if the CMB measurements were misled by some new physics,
e.g.~a new undetected relativistic particle, then the misinterpretation
could potentially spread to the low-redshift application and bias
the inferences.

Here, we show that the interpretation of the low-redshift acoustic
oscillations are robust if the CMB correctly tells us the baryon-to-photon
ratio and the epoch of matter-radiation equality.
These quantities are robustly measured in the CMB.
The actual densities of matter and radiation drop out of the calculation;
only their ratio matters.
The result is that even if the physical matter density
$\rho_{\rm m}\propto\omega_{\rm m}\equiv\Omega_{\rm m}h^2$
is misinterpreted from the CMB due to undetected relativistic components,
the inferences for dark energy from the combined CMB and 
low-redshift survey data sets
are unchanged.
Knowledge of actual densities, e.g.~$\omega_{\rm m}$, translates into 
improved constraints on the
Hubble constant, $H_0=100\,h\,{\rm km}{\rm s}^{-1}{\rm Mpc}^{-1}$.

\section{The physical scale}

The acoustic peak method depends upon measuring the sound horizon, which is
the comoving distance that a sound wave can travel between the end
of inflation and the epoch of recombination \cite{HuDod}.  Nearly all of this
distance is accumulated just prior to the epoch of recombination
at $z\simeq 1100$.
The sound horizon integral depends only on the Hubble parameter $H(z)$
and the sound speed $c_s$ in the baryon-photon plasma.
If we assume dark energy is sub-dominant at $z\sim 10^3$, then
\begin{equation}
  H(z)\simeq H_0\sqrt{\Omega_{\rm m}(1+z)^3 + \Omega_{\rm r}(1+z)^4} =
  \sqrt{\Omega_{\rm m}H_0^2(1+z)^3}\sqrt{1+{1+z\over1+z_{\rm eq}}},
\end{equation}
where $z_{\rm eq} = \Omega_{\rm m}/\Omega_{\rm r}$ is epoch of
matter-radiation equality.
The sound speed depends only on the baryon-to-photon ratio and
is $c_s = c/\sqrt{3(1+R)}$ with 
$R \equiv 3\rho_b/4\rho_\gamma \propto \omega_b (1+z)^{-1}$.
These two produce the sound horizon 
\begin{equation}
  s = \int_0^{t_{\rm rec}}\,c_s\, (1+z)dt
    = \int_{z_{\rm rec}}^\infty {c_s\;dz\over H(z)}
    = {1\over\sqrt{\Omega_m H_0^2}}
    {2c\over \sqrt{3z_{\rm eq}R_{\rm eq}}}
      \ln { \sqrt{ 1 + R_{\rm rec}} + \sqrt{ R_{\rm rec} + R_{\rm eq}}
      \over 1 + \sqrt{R_{\rm eq}} },
\label{eq:rsound}
\end{equation}
where `rec' and `eq' refer to recombination and  equality respectively.
One sees that the aside from a prefactor of $\omega_{\rm m}^{-1/2}$, the 
sound horizon depends only on the baryon-to-photon ratio and the
redshift of equality.

The epoch of recombination, being controlled by atomic physics, is very
insensitive to the cosmology.  For reasonable changes in the early universe
and our current uncertainties of the theory of recombination \cite{RecFast},
any shift in $z_{\rm rec}$ is negligible.
The baryon-to-photon ratio is also exquisitely well measured in the CMB
power spectrum by both the ratios of the odd and even acoustic peaks
and by the Silk damping tail \cite{Signatures,DampingTail}.
The former effect depends only on the gravitational inertia of the baryons
driven by the potentials near the epoch of recombination.
Thus the modulation gives us a precise measurement of the baryon-to-photon
ratio $\rho_{\rm b}/\rho_\gamma$, which with our precise knowledge of
$T_{\rm cmb}$ fixes
$\rho_{\rm b}\propto\omega_{\rm b}\equiv\Omega_{\rm b}h^2$.
Moreover, for the established value of $R\simeq 0.6$ near $z\simeq 10^3$,
the effect on the sound horizon is already small.
It seems very likely that the CMB will determine the baryon-to-photon
ratio to sufficient accuracy for this portion of the sound horizon
computation \cite{HETW}.

Information about matter-radiation equality is encoded in the amplitudes
of the peaks through the manner in which the potentials evolve as they
cross the horizon: the potential envelope \cite{Signatures,DampingTail}.
Measurements of the potential envelope thus robustly constrain equality.
Normally, one interprets this constraint as the matter density
$\omega_{\rm m}$,
on the assumption that the photons and standard neutrino background are
the full radiation density.  However, one could imagine other relativistic
components, and in this case, measuring the redshift of equality does not
imply the matter density $\omega_{\rm m}$ (we continue to assume that
the extra components are ``undetected'' in the CMB and return to this
point in the next section).
As we can see from Eq.~(\ref{eq:rsound}), the dependence of $s$ on
$z_{\rm eq}$ is relatively small since $z_{\rm eq}\simeq 3z_{\rm rec}$,
thus even a crude determination suffices to fix $s$ up to an overall
factor of $\omega_{\rm m}^{-1/2}$, i.e., $\omega_{\rm m}^{1/2}s$ is very
well measured.  The sound horizon decreases by only 
5\% if $z_{\rm eq}$ is lowered by $500$!

Understanding the acoustic oscillations at $z\simeq 1100$ allows us
to translate knowledge of the sound horizon into knowledge of wavelength
of the baryonic features in the mass power spectrum up to the same
normalization uncertainty.
We then wish to consider the measurement of this scale at lower redshift,
such as could be accomplished in large galaxy surveys.  Measuring
the scale along and across the line of sight, as a redshift or angular
scale, constrains $H(z)^{-1}/s$ and $d_A/s$, respectively.  
For dark energy with (constant) equation of state $w$, the low-redshift
quantities can be written as
\begin{equation}\label{eq:H}
H(z)^{-1} = \omega_{\rm m}^{-1/2}\left[ (1+z)^3 +
  (\Omega_{\rm de}/\Omega_{\rm m})(1+z)^{3+3w}
\right]^{-1/2},
\end{equation}
and (for zero curvature)
\begin{equation}\label{eq:Da}
  d_A(z)\propto \int_0^z {dz\over H(z)}
  \propto \omega_{\rm m}^{-1/2} \int_0^z {dz\over (1+z)^{3/2} }
  \left[1+\left(\Omega_{\rm de}/\Omega_{\rm m}\right)(1+z)^{3w}\right]^{-1/2}.
\end{equation}
Because $s\sqrt{\omega_{\rm m}}$ is well constrained, we find that
the observations actually constrain $\sqrt{\omega_{\rm m}}d_A$
and $\sqrt{\omega_{\rm m}}H(z)^{-1}$ \cite{Lin}, which contain only
the terms that depend on the bare $\Omega$ values \footnote{%
The same thing occurs in non-flat cosmologies.  For example, in an open
cosmology, we have $d_A = R_c\sinh(r/R_c)$, where $r$ is equation 
\protect\ref{eq:Da}.  But we can write
$R_c = c\sqrt{\Omega_{\rm m}\Omega_{\rm k}}/\sqrt{\Omega_{\rm m} H_0^2}$,
where $\Omega_k$ is the usual curvature term.  This substitution shows
that $\omega_{\rm m}^{1/2}d_A$ depends only on the bare $\Omega$'s.},
i.e., $\Omega_{\rm m}$, $\Omega_{\rm de}$, etc.
In other words, the prefactors of $\omega_{\rm m}^{-1/2}$ have canceled
out between $s$ and the low-redshift distances.
We can thus reliably predict the distance {\it ratios\/} between low $z$
and $z\simeq 1100$ as a function of redshift and hence constrain cosmology.

What has happened is simply that the overall scale of the universe
doesn't affect any of the distance ratios.  Usually this scale is
labeled as the Hubble constant, such that $H_0$ drops out.  We have
rephrased the scale as $\omega_{\rm m}$, despite the fact that
this would appear to include $\Omega_{\rm m}$ in addition to $H_0$.
Another way of saying this is that the standard ruler defined by the 
CMB at constant redshift of equality actually scales as $1/H_0\sqrt{\Omega_m}$.
Hence, even $z=0$ redshift-space measurements of this ruler do not 
measure $H_0$, but instead measure $\Omega_m$.

Parameter estimation for acoustic oscillations based on standard models
\cite{BlaGla,Lin,HuHai,SeoEis}
will be unchanged by the presence of undetectable new radiation provided
that the redshift of equality is measured correctly.
Only the Hubble constant would be incorrect.
In the context of CMB parameter estimation, it would be more useful
to report $s\sqrt{\omega_m}$ rather than $s$ itself, as this is the 
quantity that isolates the cosmological densities at low redshift.

Massive neutrinos should be counted as radiation at $z>10^3$ but would
be counted as matter at low redshift.  That does create a small error
in the dark energy inferences.  One would be computing $\omega_{\rm cdm}^{1/2}s$
from the CMB, but the low-redshift distances in equations (\ref{eq:H}) and (\ref{eq:Da})
would still have prefactors of
$\omega_{\rm m}^{-1/2} = (\omega_{\rm cdm}+\omega_{\rm hdm})^{-1/2}$.
As the observations constrain $d_A/s$ and $H(z)^{-1}/s$, 
the evaluation of these quantities 
in terms of bare $\Omega$'s will be altered by 
$(1+\Omega_{\rm hdm}/\Omega_{\rm cdm})^{-1/2}$.
For the massive neutrino fraction inferred for
non-degenerate neutrino species and the atmospheric neutrino mass
splitting \cite{AtmoNu} this is a 0.2\% correction; however, it could be a
few percent correction at the upper limit of the allowed masses for
degenerate neutrino species \cite{NuMass}.  Fortunately, such neutrino masses 
should be detectable in upcoming data due to their suppression of the 
late-time matter power spectrum \cite{FisherNu}.

Because the integral for the sound horizon extends to early times
(essentially to the time when the cosmic perturbations were established),
alterations to the Hubble parameter (eq.~[\ref{eq:H}]) could alter
the sound horizon even at fixed redshift of equality.  For example,
there might exist a non-relativistic particle that decays into
relativistic unseen particles sometime prior to recombination
\cite{BarBonEfs,ChuKimKim,WhiGelSil,BHMS}.
However, such decays create alterations to the gravitational potentials 
that would be detectable in the CMB if the horizon scale at that epoch
is visible in the primary anisotropies 
(i.e., wavenumbers $k\lesssim0.2h{\rm\;Mpc^{-1}}$).
This makes decays at $10^3\lesssim z\lesssim 10^5$ difficult to hide.
Furthermore, such decays alter the transfer function and would affect the
amplitude of the late-time matter power spectrum.  Precision measurement 
of the matter power spectrum out to $k\sim1h{\rm\;Mpc^{-1}}$ from the 
Lyman-alpha forest could push the limits on particle decays to earlier times.
In the standard theory, the first 1\% of the sound horizon integral
is contributed by $z=200,000$.  It therefore seems very unlikely that
a decaying particle could significantly affect the sound horizon 
(relative to the standard theory at fixed redshift of equality) and escape
detection in the CMB or large-scale structure.

As a corollary, it is interesting to note that when one measures the
acoustic oscillation scale at $z_{\rm rec}$ with the CMB and at some low
redshift $z_g$ in a galaxy survey, one can construct an observable (the
difference of the suitably scaled angular wavenumbers of the acoustic
peaks) that isolates the cosmology between $z_g$ and $z_{\rm rec}$, i.e.,
\begin{equation}
\ell_{z_{\rm rec}} - \ell_{z_{\rm g}} \propto 
\int_{z_g}^{z_{\rm rec}} {dz\over \omega_{\rm m}^{-1/2} H(z)}
\end{equation}
This integral is trivial when the universe is completely matter-dominated.
Hence, if we can perform a galaxy survey at some suitably high redshift,
e.g., $z_g=4$, where the dark energy is supposed to be negligible,
we could search for dynamical shenanigans at $4<z<10^3$.  
A photometric-redshift survey over large amount of sky would be an 
economical route to this goal, as one can acquire very large
survey volumes \cite{SeoEis}.

\section{Whither the actual density?}

\begin{figure}[tb]
\begin{center}
\resizebox{3in}{!}{\includegraphics{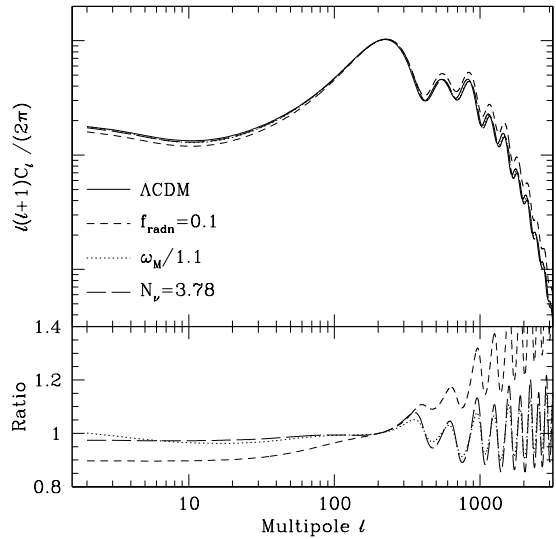}}
\end{center}
\caption{The angular power spectra for a fiducial $\Lambda$CDM model
(solid) and three variants with a lower $z_{\rm eq}$.  The dotted line
shows a model with $\omega_{\rm m}$ reduced by 10\%, the long dashed
line $N_\nu$ increased from 3.04 to 3.78 (which raises $\rho_{\rm r}$
by 10\%) and the short dashed line 10\% extra $\rho_{\rm r}$ in the
form of a fluid.  The upper panel shows the spectra while the lower
panel shows the ratio to the $\Lambda$CDM curve.  All spectra have
been normalized (arbitrarily) at $\ell=200$.}
\label{fig:cl}
\end{figure}

We have argued that only the ratios of the radiation and matter densities
enter the acoustic oscillation standard ruler method, but of course it
is interesting to measure the actual values of the densities as they
might reveal new relativistic components or unexpected evolution
\footnote{We assume implicitly that any extra radiation is introduced
between 1 second and 10,000 years, since radiation present prior to 1 second
would modify big bang nucleosynthesis.}.

The CMB is sensitive to the non-photon radiation density primarily through
its effect on equality and the evolution of the potentials.  This encodes
a sensitivity to both the amount of radiation and its quadrupole moment.
In general any particle that free streams rather than behaves as a fluid
will have a local quadrupole that will affect the evolution of the potentials
and thus the anisotropy in the CMB.
This breaks the degeneracy in the CMB between changes in $\omega_{\rm m}$
and $\omega_{\rm rad}$ \cite{HSSW} and allows us to constrain a
``conventional'' neutrino at the $N_\nu\simeq 0.1-0.2$ level \cite{HETW}.
At the other extreme, extra radiation which is of the perfect fluid form
leads to an increase in the small-scale power which is easily discerned from
a change in $\omega_{\rm m}$ (see Fig.~\ref{fig:cl}).
Only a radiation component whose higher moments track closely that of the
traditional mix of photons and neutrinos could be confused with a change
in $\omega_{\rm m}$.

Altering the radiation and matter density while holding the baryon
(and photon) densities fixed would alter the baryon fraction $\Omega_b/\Omega_m$
and would produce significant offsets in the late-time matter power spectrum 
(e.g.~\cite{BE84,Hol,EH98}).
Higher baryon fractions suppress power on small scales compared to large,
with the transition occurring near the sound horizon at
$k\approx0.05h{\rm\;Mpc^{-1}}$.  
The acoustic oscillations shortward of this scale would also be increased.
The small-scale amplitude (e.g., $\sigma_8$) would be reduced.  These
changes are observable in galaxy surveys, weak lensing surveys, and cluster
abundance measurements.
Of course one can also directly measure the Hubble constant \cite{KeyProject}.

In short, although the acoustic peak method depends only on the redshift
of equality, other cosmological measurements both at $z=1000$ and at $z\sim0$
are well poised to measure the actual densities and thereby constrain the 
presence of unknown relativistic species.

\section{Conclusions}

We have shown that the standard ruler defined by the acoustic oscillations
prior to recombination can deal gracefully with uncertainties in the 
matter density $\omega_m$ provided that the redshift of matter-radiation
equality is well measured.  The anisotropies of the CMB have very
good leverage on this quantity, and so the acoustic peak method of
probing dark energy is robust.  In addition, it is likely that the CMB, perhaps
in combination with other probes, will be able to constrain the actual
matter and radiation densities.

\medskip
Acknowledgements: We thank Eric Linder and Nikhil Padmanabhan for useful
discussions and the Lawrence Berkeley
Laboratory for the hospitality of its summer workshop program where
this work was begun.  
DJE is supported by National Science Foundation grant AST-0098577
and by an Alfred P.~Sloan Research Fellowship.
MW is supported by the NSF and NASA.

\end{document}